\documentstyle[amssymb,preprint,aps]{revtex}

\tightenlines

\begin{document}
\title{Critical current of a S/F-I-N/S tunnel structure\\
in a parallel magnetic field}
\author{V. N. Krivoruchko, E. A. Koshina}
\address{Donetsk Physics \& Technology Institute, Donetsk , Ukraine}
\date{\today}
\maketitle
\pacs{74.50.+r, 74.78.-w, 74.45.+c}

\begin{abstract}
We study the critical current of a S/F-I-N/S tunnel structure - a proximity
coupled superconductor (S) with thin ferromagnetic (F) and nonmagnetic (N)
metals separated by an insulating (I) barrier - in an external magnetic
field. The dependence of the current on exchange and external magnetic
fields, and their mutual orientation is analyzed within the microscopic
theory of the proximity effect. We find that the S/F-I-N/S contact with
strong magnetism of the F layer is in a ground state with superconducting
phase difference on the electrodes about $\pi /2$ . Low external magnetic
field reverses the critical current sign of such a contact (a $\pi $-state
of the junction) for parallel field orientation. For antiparallel alignment
the dc Josephson current behavior is essentially nonmonotonic: by increasing
an external magnetic field the current passes though a maximum, while the
0-phase state is held, and then the junction gets into the $\pi $-phase
state with an opposite current direction. We demonstrate that the hybrid
S/F-I-N/S metal structures possess new effects of the superconductivity in
the presence of spin splitting, while an experimental set-up seems to be
feasible and it may lead to further understanding of superconducting
proximity effects in ferromagnetic materials.
\end{abstract}

\section{Introduction}

The problem of superconductivity in the presence of spin splitting is
currently heavily investigated. In fact, this question was already addressed
experimentally and theoretically a while ago for the simplest case of a very
thin superconducting film in a parallel magnetic field. For such geometry
the quasiparticle energy states in a superconductor are splitted due to the
interaction of the field with spin magnetic moments of the quasiparticles,
while the orbital effects are negligible. This splitting has been earlier
demonstrated by extensive studies of tunnel junctions formed by thin
superconducting films (see, e.g., [1-3] and references therein).

A more sophisticated case is the proximity effected hybrid systems
consisting of a superconductor (S) and a ferromagnet (F). In a ferromagnet,
the difference in the energy between spin-up and spin-down bands is an
internal property, and that is the physical reason why a singlet Cooper
pair, injected from a superconductor into a ferromagnet, acquires a finite
momentum. The resulting proximity induced superconducting state of the F
layer is qualitatively different from the zero-momentum state: it is
spatially inhomogeneous and the order parameter contains nodes where the
phase changes by $\pi $ (see recent review [4] and references therein).
Particularly, transport properties of the S/F structures have turned out to
be quite unusual. Experiments that have been performed on SFS weak links
[5-7] and SIFS\ tunnel junctions [8]\ directly prove the so-called $\pi $%
-phase superconductivity. Planar tunneling spectroscopy also reveals a $\pi $%
-phase shift in the order parameter, when superconducting correlations
coexist with the ferromagnetic order [9]. An inverse phenomenon, the
modification of the BCS density of states in mesoscopic S layer under the
influence of a ferromagnet is observed in tunnel spectroscopy experiment
[10] as well. For S/F-I-F/S tunnel structures with very thin F layers, it
was also predicted [11,12] that one can, if there is a parallel orientation
of the F layers magnetization, turn the junction into the $\pi $-phase state
with the critical current inversion, or, if there is an antiparallel
orientation of the F layers internal fields, save the 0-phase junction state
but enhance the tunnel current.\ Experimentally, however, the enhancement of
the dc Josephson current has not been detected until now. One of the main
problems here is the requirement to fix one layer magnetization of the
sample, while preserving a possibility to reorient the magnetization of
another layer by a weak magnetic field.

In this report we draw attention to another experimental set-up where, as we
show below, there is the critical current enhancement, as well as other new
effects of the superconductivity in the presence of spin splitting, while
the experimental set-up admits of simple variations of an applied field
direction. Namely, we consider the dc Josephson effect for tunnel structure
where there is spin splitting for two reasons - due to an external magnetic
field and due to an internal exchange field of the F layer. The ground state
and critical current dependence on the applied field is analyzed for a
system of proximity coupled bulk superconductor with thin ferromagnetic and
thin normal nonmagnetic metals separated by an insulating barrier (S/F-I-N/S
tunnel junction). Based on a microscopic theory of the superconducting
proximity effect for S/F and S/N bilayers we demonstrate that the S/F-I-N/S
contact with strong enough magnetism of the F layer is in a ground state
with $\pm \pi /2$ superconducting phase difference on the electrodes. In
such a contact low external magnetic field reverses the critical current
sign (the $\pi $-state of the junction) if the fields (external and
internal) have parallel orientation. For the antiparallel fields alignment
the critical current behavior is much more complicated: a monotonic increase
of the external field causes an enhancement of the critical current, which
is then changed by a reverse of current sign.

The article is organized as follows. In Sec. II we outline the model of the
junction. In Sec. III the outcomes of the theory of proximity coupled S/F
and S/N bilayers have been generalized to include external magnetic field
effects. The main results of the paper are presented in Sec.IV. In Sec. V we
evaluate the approximations that have been made in the theory and briefly
discuss a possible realization of the S/F-I-N/S systems to explore the
predicted effects.

\section{The model of a junction}

The system we are interested in is the S/F-I-N/S tunnel structure in a
parallel magnetic field shown in Fig.1. We assume that the first (left)
electrode of the junction is formed by a proximity coupled S/F bilayer while
the second (right) is formed by a proximity coupled S/N bilayer. The
transparency of the insulating layer is small enough to neglect the effect
of a tunnel current on the superconducting state of the electrodes. The S/F
and S/N interfaces may have arbitrary finite transparency, but it is large
compared to the transparency of the junction barrier. The F layer will be
treated as a single domain ferromagnet. We also expect the F layer
magnetization to be aligned parallel to the interface, so that it does not
create a spontaneous magnetic flux penetrating into the S layer. The general
relation between\ the Curie temperature of the F metal and the
superconducting transition temperature of the S metal is considered.
Transverse dimensions of the junction are supposed to be much less then the
Josephson penetration depth, so that (a) all quantities depend only on a
single coordinate $x$ normal to the interface surface of the materials, and
(b) a flux quantum can not be trapped by the junction.

We assume that the applied field penetrates and mostly concentrates into the
''weak'' F-I-N part of the junction. We will also neglect the effect of the
magnetic field on electron orbits as compared with the effect of the field
on electron spins. This assumption is well satisfied for films thinner than
their superconducting coherence lengths $d_{F(N)}<<\xi _{F(N)}$ . Here $%
d_{F,N}$\ is the thickness of the F and N layers, respectively; $\xi _{F}%
\symbol{126}\sqrt{D_{F}/2H_{e}}$ and $\xi _{N}\symbol{126}\sqrt{D_{N}/2\pi
T_{C}}$ are superconducting coherence lengths for F and N metals; $D_{F,N}$\
is a diffusion coefficient in the F, N metal; $H_{e}$ is the exchange
energy. For films of such thicknesses one can also be sure that there is a
fairly uniform penetration of a parallel field.\ (The evaluation of these
approximations will be made below, in Sec. V). We consider the ''dirty''
limit for all metals, i. e., we assume that $l_{S}<<\xi _{S}$ \ and $%
l_{F(N)}<d_{F(N)}$, and that the condition $H_{e}<\tau ^{-1}$\ is fulfilled
[13], where $l_{S,F,N}$ is the electron mean free path for the S, F, and N
layers, respectively; $\xi _{S}$ is the superconducting coherence length in
the S metal; and $\tau $ is the momentum relaxation time. (We have taken the
system of units with $\hbar =\mu _{B}=k_{B}=1$). We restrict ourselves to
the case when the spin-orbit scattering is absent and spin ''up'' and spin
''down'' electron subbands do not mix with each other. In equations below,
we will use the effective coherence length $\xi \symbol{126}\sqrt{D_{F}/2\pi
T_{C}}$ for the F layer, thus providing the regular crossover to both limits 
$T_{C}>>H_{e}\longrightarrow 0$ and $H_{e}>>T_{C}$ (see Refs. 14, 15 for
details); the relation on film thickness is read as $d_{F}<<\xi _{F}$ or $%
\xi $ , depending on the relation between $T_{C}$ and $H_{e}$.

\section{S/F and S/N sandwiches in the magnetic field}

As a first step, we should generalize results of the theory for a proximity
coupled S/F (S/N) bilayer of a massive, $\xi _{S}<<d_{S}$ , superconductor
and a thin ferromagnetic (nonmagnetic) metal taking into account the
influence of the applied magnetic field $H$ . The most direct way to do it
is to use the results for a S/F proximity coupled bilayer (Refs. [12,14,15]).

As is well known, the properties of ''dirty'' metals can be described by the
Usadel equations [16]. Following [17], we introduce the modified Usadel
functions $\Phi $ , used to take into account the normalized confinement on
the normal $G$ and anomalous $F$ Green's functions. Due to small thickness
of the normal (magnetic and nonmagnetic) metal, the proximity effect problem
can be reduced to a boundary problem for the S layer and a relation for
determining modified Usadel function $\Phi $\ of a nonsuperconducting layer
(see Refs. [14,15,17,18] for details).\ The boundary conditions at the SF
(SN) interface are modelled by two parameters: $\gamma _{M}$ and $\gamma
_{B} $ . The $\gamma _{M}$ plays the role of an effective pairbreaking
parameter near the SF (SN) interface and its value is mainly determined by
electron densities of S and F (N) metals. A large value of $\gamma _{M}$
corresponds to high density of quasiparticles in the F (N) metal compared to
that in the S near the SF (SN) boundary. In this case the diffusion of
quasiparticles into the superconductor leads to a strong suppression of the
order parameter in the S region. The scale at which superconductivity
reaches the bulk value is $\xi _{S}$ from the SF (SN) interface. In the
opposite case, $\gamma _{M}<<1$ , the influence of the F (N) layer on the
superconductivity of the S metal is weak, and even vanishes if $\gamma
_{M}=0 $. The parameter $\gamma _{B}$ $\symbol{126}$ $R_{B}$ , where\ $R_{B}$
is the SF (SN) boundary resistance, describes the electrical quality of the
SF (SN) interface. The case $\gamma _{B}<<1$ corresponds to a vanishing
interlayer resistance, i.e., the S and F (N) metals are in good electric
contact. The case $\gamma _{B}>>1$ corresponds to low transparency of the SF
(SN) barrier. There is also one more parameter, $H_{e}$ , that is the value
of the exchange spin splitting in the F layer. In the general case, the
problem needs self-consistent numerical calculations of the Usadel
equations. However, at some simplifying assumptions, the model admits of
analytic solutions that hold all new physics we are interested in.

For F (N) films thinner than their superconducting coherence lengths, $%
d_{F(N)}<<\xi _{F(N)}$ , the critical current of the S/F-I-N/S tunnel
contact can be represented through the S layer Green's functions (see Eq.
(3), below). The explicit expressions for these functions can be obtained on
the basis of the microscopic theory of proximity effects for S/F and S/N
bilayers for a weak, $\gamma _{M}<<1$\ , and a strong, $\gamma _{M}>>1$\ ,
proximity effect. We generalize these results assuming the external magnetic
field penetrates into the F and N layers. Namely, in the case of a bulk S
layer proximity coupled with a thin F layer with a weak proximity effect $%
\gamma _{M}<<1$ and an arbitrary value of the boundary resistance $\gamma
_{B}$ (quite a realistic experimental case) for the function $\Phi _{SF}$ we
have the analytical solution (see Eq. (14), Ref. 14 or Eq. (10), Ref. 15): 
\begin{equation}
\Phi _{SF}(\omega )\equiv \Phi _{SF}(\omega ,x=0)=\Delta _{S}\left( 1-%
{\displaystyle{\gamma _{M}\beta \omega _{F} \over \gamma _{M}\beta \omega _{F}+\omega A_{F}}}%
\right) \text{ \ ,}
\end{equation}
where $\Delta _{S}$ is the absolute value of the superconducting order
parameter in the bulk of the S layer, $\omega _{F}=\omega +i\sigma
(H_{e}+H), $ $\omega \equiv \omega _{n}=\pi T(2n+1),n=0,\pm 1,\pm 2,...$ are
Matsubara frequencies , $\sigma =\pm 1$ is the spin sign;$\ \ \beta
^{2}=\left( \omega ^{2}+\Delta _{S}^{2}\right) ^{1/2}/\pi T_{C},$ and\ $%
A_{F}=\left[ 1+\gamma _{B}\omega _{F}\left( \gamma _{B}\omega _{F}+2\omega
/\beta ^{2}\right) /(\pi T_{C})^{2}\right] ^{1/2}$. The sign of the external
magnetic field depends on the relative orientation to the internal exchange
field. To be definite, we will suppose that $H_{e}>0$ and there will be the
case of the parallel external fields orientation if $H>0$, while for $H<0$
the fields will have antiparallel orientation.

For the S/F bilayer with a strong proximity effect $\gamma _{M}>>1$ and free
value of the boundary resistance $\gamma _{B}$ our calculations yield (see
Eq. (12) Ref. 15): 
\begin{equation}
\Phi _{SF}(\omega ,x=0)\equiv \Phi _{SF}(\omega )=B(T)\left( \pi
T_{C}+\gamma _{B}\omega _{F}\right) /\gamma _{M}\omega _{F}
\end{equation}
where $B(T)=2T_{C}[1-(T/T_{C})^{2}][7\zeta (3)]^{-1/2}$ and $\zeta (3)$ is
the Riemann $\zeta $ function.

Qualitatively similar results are obtained for the S/N bilayer. If the
influence of the N layer on the superconductivity of the S metal is weak, $%
\gamma _{M}<<1$ and the value of interface resistance is arbitrary, for $%
\Phi _{SN}(\omega )$ we have the expression (1) where the substitutions
should be made: $\omega _{F}\longrightarrow \omega _{N}=\omega +i\sigma H$
and $A_{F}\longrightarrow A_{N}=\left[ 1+\gamma _{B}\omega _{N}\left( \gamma
_{B}\omega _{N}+2\omega /\beta ^{2}\right) /(\pi T_{C})^{2}\right] ^{1/2}$.
In the opposite case of a strong proximity effect $\gamma _{M}>>1$ , for $%
\Phi _{SN}(\omega )$ we have the relation (2) where $\omega _{F}$ is changed
by $\omega _{N}$ , and, as before, there is an arbitrary value of the SN
boundary resistance.

Below we will suppose for simplicity, that the interface parameters involved
, $\gamma _{M}$ and $\gamma _{B}$ , have the same values for both SF and SN
interfaces. An extension to different SN and SF boundary parameters is
straightforward, and will be made towards the end of the Sec. IV. An
additional physical approximation is also the assumption that the spin
discrimination by the interfaces is unimportant. Generalization for
interfaces with different transmission probabilities for quasiparticles is
obvious and we do not present it here.

\section{Critical current}

The critical current of a S/F-I-N/S tunnel contact can be represented in the
form (see, e.g., Ref. [15]) 
\begin{equation}
j_{C}=(eR_{N}/2\pi T_{C})I_{C}=(T/T_{C})%
\mathop{\rm Re}%
\sum_{\omega >0,\sigma }\{G_{S}\Phi _{S}/\omega \}|_{F}\{G_{S}\Phi
_{S}/\omega \}|_{N}\times
\end{equation}
\[
\{[1+2\omega _{F}G_{S}(\gamma _{B}/\pi T_{C})+\omega _{F}^{2}(\gamma
_{B}/\pi T_{C})^{2}]_{F}\times \lbrack 1+2\omega _{N}G_{S}(\gamma _{B}/\pi
T_{C})+\omega _{N}^{2}(\gamma _{B}/\pi T_{C})^{2}]_{N}\}^{-1/2} 
\]
where R$_{N}$ is the resistance of the contact in the normal state; the
subscript F (N) labels quantities referring to the left ferromagnetic and
right nonmagnetic electrodes. Here we used $G_{S}=\omega /(\omega ^{2}+\Phi
_{S}\widetilde{\Phi }_{S})^{1/2}$ , where $\widetilde{\Phi }_{S}(\omega
,H_{eff})=\Phi _{S}^{\ast }(\omega ,-H_{eff})$ and $H_{eff}$ is $(H+H_{e})$\
or $H$ .

\subsection{Strong proximity effect}

Let us first present results for the case of a strong proximity effect, $%
\gamma _{M}>>1$ , and a vanishing interface resistance, $\gamma _{B}=0$. If
there is a strong suppression of the order parameter near S/F (S/N) boundary
then, using the expressions for $\Phi _{SF}$ and $\Phi _{SN}$ , we can
recast the Eq. (3) in the form: 
\begin{equation}
j_{C}(H,H_{e})\approx 2(T/T_{C})B_{M}^{2}(T)\sum_{\omega }\frac{\omega
^{2}-H(H_{e}+H)}{\omega ^{2}\{[(\omega ^{2}-H(H_{e}+H)]^{2}+\omega
^{2}(H_{e}+2H)^{2}\}}
\end{equation}
where $B_{M}(T)=B(T)\pi T_{C}/\gamma _{M}$\ , and in proceeding these
relations, we have taken into account that the value of $\Phi _{SF(SN)}$ is
small, $\Phi _{SF(SN)}\symbol{126}\gamma _{M}^{-1}<<1$. Performing the
substitution $(H_{e}+H)\rightarrow H_{eR}$ and $H\rightarrow \pm H_{eL}$\ in
Eq. (4) , one can obtain the results for the (S/F)$_{L}$-I-(F/S)$_{R}$
tunnel structure for (anti)parallel orientation of the $H_{eR}$ and $H_{eL}$
\ fields (see Eq. (15) in Ref. 15 and Eq. (5) in Ref. 12). If $H_{e}=0$ \
and $H=0$ one can also restore the result for S/N-I-N/S junction (see, e.g.,
Eq. (31) of Ref. 17 and Eq. (46) of Ref. 18). In the general case, as it is
seen from Eq. (4), in\ the geometry with the parallel fields by increasing
the external magnetic field one can inverse the supercurrent sign of the
S/F-I-N/S tunnel structure. In other words, at large enough applied field, $%
H(H_{e}+H)>>\omega ^{2}\symbol{126}(\pi T_{C})^{2}$\ , the critical current
changes its sign in comparison with that when $H\rightarrow 0$ . In
particular, from Eq. (4) we obtain in the limit of large field values: 
\[
j_{C}(H,H_{e})\text{ \ }\symbol{126}-B^{2}(T)/\gamma _{M}^{2}\sum_{\omega
}1/\{\omega ^{2}H(H_{e}+H)\}\text{ \ }\symbol{126}\text{ }%
-(T_{C}-T)^{2}/\{\gamma _{M}^{2}H(H_{e}+H)\}, 
\]
i. e. , a crossover of the S/F-I-N/S junction from the 0-phase state to the $%
\pi $-phase state takes place.

The most interesting results have been obtained for the geometry with the
opposite field orientation: $H<0$. Simple substitution into Eq. (4) gives
for $H=-H_{e}$ that $j_{C}(-H_{e},H_{e})=j_{C}(0,H_{e})$ , while if $%
H=-H_{e}/2$ we directly obtain that $j_{C}(-H_{e}/2,H_{e})>j_{C}(0,H_{e}),$ $%
j_{C}(-H_{e},H_{e})$ . Comparing, we also find that $j_{C}(-H_{e}/2,H_{e})%
\neq j_{C}(H_{e}/2,H_{e})$ and $j_{C}(H_{e},H_{e})\neq j_{C}(-H_{e},H_{e})$
. As it follows from Eq. (4), for large enough values of the external field,
the critical current again inverses its sign and the crossover of the
S/F-I-N/S junction from the 0-phase state to the $\pi $-phase state takes
place. However, while increasing the field in the interval $0<|H|<H_{e}$ , a
nonmonotonic current versus field dependence is observed. The main features
of this dependence are shown on Fig. 2, where the Josephson current
amplitude of the S/F-I-N/S junction versus the applied magnetic field is
given for the case of a strong proximity effect, $\gamma _{M}=10$ and $%
\gamma _{B}=0$ , and different exchange field intensity. (Here and below,
dashed curves are used for the results of the numerical calculations for the
parallel fields orientation and full curves are used for the results in the
opposite geometry). It can be seen, that the dc current greatly depends on
relative orientation of the fields.

To reveal the underlying physics behind the critical current behavior on
field, let us consider a ground state and a phase shift across the contact
at zero external field. As follows from Eq. (2), the modified Usadel
function $\Phi _{SF}(\omega )$ can be written in the form (the case $\gamma
_{B}=0$) 
\[
\Phi _{SF}(\omega )=B(T)(\pi T_{C}/\gamma _{M})\frac{\omega -iH_{e}}{\omega
^{2}+H_{e}^{2}} 
\]
Taking into account that the typical value $\omega $ \symbol{126}$\pi T_{C}$
, one can see that in the limit $H_{e}>>\pi T_{C}$ the unified correlation
function of the S/F bilayer acquires an additional $\pm \pi /2$ phase shift
in comparison with the similar function for the S/N bilayer. When the
external field is applied, $H\neq 0$ and $\gamma _{M}>>1$ , $H_{e}>>\pi
T_{C} $ , we may evaluate the critical current as 
\begin{equation}
j_{C}\text{ }\symbol{126}\text{ }%
\mathop{\rm Re}%
\sum_{\omega }F_{SF}(\omega )F_{SN}(\omega )\text{ \symbol{126}}%
\mathop{\rm Re}%
\sum_{\omega }\Phi _{SF}(\omega )\Phi _{SN}(\omega )/\omega ^{2}\text{ 
\symbol{126}}
\end{equation}
\[
\text{ \symbol{126}}%
\mathop{\rm Re}%
\sum_{\omega }\frac{\exp \{\pm i[\pi /2+\arctan (H/\omega )]\}}{\omega
^{2}(\omega ^{2}+H^{2})^{1/2}} 
\]
i.e., for $H\longrightarrow 0$\ ,\ the value of the supercurrent may be low
enough - the exchange field blocks a contact. So, the S/F-I-N/S contact with
strong enough magnetism of the F layer can have a ground state with the
superconducting phase difference on the electrodes about $\pi /2$. Note,
that this mechanism of the $\pi /2$ junction\ realization should not be
confused with the $\pi /2$-phase state induced by current fluctuations in
the junction plane, the situation that has been treated earlier in Ref. [19]
for SFS structures with a thick F layer. We postpone a more detailed
description of the $\pi /2$ state of S/F-I-F/S and S/F-I-N/S hybrid\ systems
to a forthcoming publication [20], but here discuss the effects of the
external magnetic field. As one can see from expression (5), if the contact
is in magnetic field, the additional phase shift of the\ Usadel function $%
\Phi _{SN}(\omega )$ has been induced. Depending on the relative
orientation, the applied field can prompt the total phase shift across the
contact to a $\pi $-state of the junction or to a $0$-phase state. For the
latter case one can expect an essential enhancement of the critical current
amplitude. Such type of behavior we can see on Figs. 2-7.

As seen on Fig. 2, a low external magnetic field reverses the critical
current sign if the fields (external and internal) have parallel
orientation. For antiparallel fields alignment the critical current behavior
is much more complicated. In some interval of the applied field the
enhancement of the dc Josephson current takes place in comparison with the
case of $H=0$, while the 0-phase state of the junction is held. By
increasing the magnetic field one can turn the current though a few extremum
and then adopt the junction to the $\pi $-phase state superconductivity with
the opposite current direction.

The results above are obtained for the case of a vanishing interface
resistance, $\gamma _{B}=0$. After simple but cumbersome algebra, one can
also obtain expressions for $\gamma _{M}>>1$ , and $\gamma _{B}\neq 0$ . We
will not extract here the formulae but illustrate our calculations on Fig.
3, where the dc current dependence versus the applied magnetic field is
shown for the S/F-I-N/S junction with a strong influence of the F (N) layer
on superconducting layers , $\gamma _{M}=10$ , and different interface
transparency: $\gamma _{B}=1\div 10$. A critical current amplitude of the
junction depends greatly on bilayers parameters, and on the interface
transparency $\gamma _{B}$ , in particular, decreasing if the transparency
decreases,\ as it is seen on Fig. 3.

\subsection{Weak proximity effect}

Using the expression (1), after simple transformations one can obtain for
the dc Josephson current in the case of vanishing interface resistance, $%
\gamma _{B}=0$ , and a weak proximity effect, $\gamma _{M}<<1$: 
\begin{equation}
j_{C}(H,H_{e})\approx 2(T/T_{C})%
\mathop{\rm Re}%
\sum_{\omega }\frac{\Delta _{S}^{2}}{\Omega ^{2}}\times \{1+2i\sigma \omega
\gamma _{M}\beta (H_{e}+2H)\Omega ^{-2}-
\end{equation}
\[
-\gamma _{M}^{2}\beta ^{2}[H^{2}+(H_{e}+H)^{2}+4\omega ^{2}\Omega
^{-2}H(H_{e}+H)]\Omega ^{-2}\}^{-1/2}\text{ \ , }
\]
where $\Omega ^{2}=\Delta _{S}^{2}+\omega ^{2}$\ . In proceeding these
relations, we have taken into account that $\gamma _{M}\beta $ is small, $%
\gamma _{M}\beta <<1$. Simple analysis shows that $%
j_{C}(0,H_{e})=j_{C}(-H_{e},H_{e})$ as it should be from the symmetry of the
fields geometry. If $\gamma _{M}\beta H_{e}<<1$ , one can also obtain that 
\[
j_{C}(-H_{e}/2,H_{e})\approx 2(T/T_{C})\sum_{\omega }\frac{\Delta _{S}^{2}}{%
\Omega ^{2}}\{1-(\gamma _{M}\beta H_{e})^{2}\frac{\Delta _{S}^{2}-\omega ^{2}%
}{2\Omega ^{2}}\}^{-1/2}>j_{C}(0,H_{e})
\]
The dependence of the Josephson current amplitude of S/F-I-N/S junction on
the applied magnetic field for a weak proximity effect, $\gamma _{M}=0.1$ ,
and different exchange field values of the F layer is shown on Fig. 4. As in
the case of strong proximity effect, the current greatly depends on the
relative orientation of the fields. For parallel geometry, a ground state
with a phase shift across the contact about $\pi /2$ is found at low field,
while with the field increasing the ground state is changed to the $\pi $
phase superconductivity. For the antiparallel orientation the current
behavior is essentially nonmonotonic and the increasing of the field turns
the current through a few extrema until the junction crossovers into the $%
\pi $-phase state with opposite current direction. Shown on Fig. 5. are
representative dc current dependences versus the applied magnetic field for
the junction with $\gamma _{M}=0.1$ , $H_{e}=1.5\pi T_{C}$ , and different
interface transparency. As is expected, the critical current amplitude
decreases with increasing the interface resistance.

As one can see on Figs. 2-5, for antiparallel geometry the $\pi $ phase
superconductivity takes place when the external field value is about the
internal one, $-H\gtrsim H_{e}$ . This result, far from being surprising, is
related to the quasiparticle spin splitting in the effective magnetic fields
in bilayers. Namely, the S/F-I-N/S structure in antiparallel magnetic field $%
H=-H_{e}$ has the same spin splitting as\ S/N-I-F/S one in zero magnetic
field. In other words, for $-H>H_{e}$\ we should have the critical current
versus field dependence as those we have for $H>0$ .\ That is what we see on
Figs. 2-5.

\subsection{Asymmetric bilayer interface parameters}

To illustrate the main features of\ the Josephson effect for the S/F-I-N/S
tunnel structure caused by an applied magnetic field, we took, for
simplicity, the same interface parameters for both SF and SN interfaces. As
the nonsuperconducting metals, - ferromagnetic and nonmagnetic ones, -
should be different, this approximation may be unrealizable experimentally.
A further step consists in extending the theory to a real system, where the
boundary conditions for S/F and S/N bilayers can be different.

Generalization of the results to SF and SN interfaces with different
transmission probabilities for quasiparticles is obvious. We present the
results for two main cases: (a) the structure with a strong proximity effect
for the SN boundary and a weak proximity effect for the SF boundary, and
vice versa: (b) a strong proximity effect for the SF boundary and a weak
proximity effect for the SN boundary. Let us first discuss the analytical
results for vanishing interface resistance $\ \gamma _{B}=0$ . We obtain 
\begin{equation}
j_{C}(H,H_{e})\text{ }\symbol{126}\text{ }(T/T_{C})B_{MN}(T)\sum_{\omega }%
\frac{\Delta _{S}}{\Omega (\omega ^{2}+H^{2})}\{1-\gamma _{MF}\beta
H(H_{e}+H)\Omega ^{-2}\}
\end{equation}
if $\gamma _{MN}>>1,$ and $\gamma _{MF}<<1$ ; here\ $B_{MN}(T)=B(T)\pi
T_{C}/\gamma _{MN}$ . As before, in\ the geometry with the parallel fields,
the increase of external magnetic field inverses the supercurrent direction
of the S/F-I-N/S tunnel structure at large enough applied field, $\gamma
_{MF}H(H_{e}+H)>>\omega ^{2}\symbol{126}(\pi T_{C})^{2}$\ . In the geometry
with the opposite field orientation, simple substitution into Eq. (7) gives
that $j_{C}(-H_{e},H_{e})=j_{C}(0,H_{e})$ , while at $H=-H_{e}/2$ we again
obtain that $j_{C}(-H_{e}/2,H_{e})>j_{C}(0,H_{e}),$ $j_{C}(-H_{e},H_{e})$ .
At low values of the field, $|H|<H_{e}$\ , the nonmonotonic current versus
field dependence can be observed; for large enough values of the external
field, $|H|>H_{e}$\ , the critical current inverses its sign and the
crossover of the S/F-I-N/S junction from the 0-phase state to the $\pi $%
-phase state takes place.

Qualitatively similar results are obtained for a junction with $\gamma
_{MN}<<1$ , $\gamma _{MF}>>1$ . If $\gamma _{B}=0$ our analytical
calculations give 
\begin{equation}
j_{C}(H,H_{e})\text{ }\symbol{126}\text{ }(T/T_{C})B_{MF}(T)\sum_{\omega }%
\frac{\Delta _{S}}{\Omega \lbrack \omega ^{2}+(H+H_{e})^{2}]}\{1-\gamma
_{MN}\beta H(H_{e}+H)\Omega ^{-2}\}
\end{equation}
where $B_{MF}(T)=B(T)\pi T_{C}/\gamma _{MF}$ .

More general cases, $\gamma _{B}\neq 0$ , are illustrated by numerical
calculations, shown on Figs. 6 and 7. In Fig.6 the dependence of the
Josephson current amplitude of the junction on the applied magnetic field is
given for a strong proximity effect for the SN boundary, $\gamma _{MN}>>1$ ,
a weak proximity effect for the SF boundary, $\gamma _{MF}<<1$, and
different interface transparency $\gamma _{B}=0\div 2$ . (Here and for Fig.
7 the numerical calculations have been made for simplicity for SF and SN
interfaces with the same resistivity, i. e., $\gamma _{BF}=\gamma
_{BN}=\gamma _{B}$ ). Opposite situation is illustrated by the results on
Fig. 7. As is seen, the main features of the current-field behavior are
robust enough and are held for the S/F-I-N/S junction with asymmetric
bilayer parameters, too. However, one can also see, that the current
amplitude depends on the conditions on the SF and SN interfaces.

In an experiment, the exchange and applied fields may be noncollinear. When
these is an angle $\theta $ between the exchange and applied fields
directions, the expression for the critical current can be written in the
form 
\[
j_{C}(\theta )=j_{C}^{P}\cos ^{2}(\theta /2)+j_{C}^{A}\sin ^{2}(\theta /2) 
\]
where $j_{C}^{P(A)}$ is the current for parallel (antiparallel) fields
alignment. The dependence of the critical current on the angle between the
fields is illustrated by Fig. 8 for S/F-I-N/S structure with $\gamma
_{M}=0.1 $, $\gamma _{B}=5$\ for the SF and SN boundaries and$\ H_{e}=\pi
T_{C}$.

\section{Conclusion}

The question of superconductivity in the presence of spin splitting is
currently heavily investigated, both theoretically and experimentally.
However, in spite of a larger number of theoretical predictions, most of
them have not been detected experimentally until now. There are some set-up
problems, and one of them is the requirement to fix one layer magnetization,
while preserving the possibility to reorient the magnetization of another
layer by a weak magnetic field. In this report we have analyzed\ a
superconducting tunnel structure where there is spin splitting for two
reasons: spin splitting caused by an external magnetic field and by an
internal exchange field. The advantage of such structure is the possibility
to get a convenient governing and adjustable parameter - an external
magnetic field. We demonstrate that in the contacts under consideration the
critical current can reverse its sign (the $\pi $-state of the junction) for
the parallel field orientation. For the antiparallel field alignment the
critical current behavior is much more complicated: for 
\mbox{$\vert$}%
$H|<H_{e}$ a monotonic increase of the\ external field causes an enhancement
of the critical current; further field's increase 
\mbox{$\vert$}%
$H|>H_{e}$\ , reverses the current direction. We reveal the physics behind
such critical current\ versus external field behavior. Namely, we show that
in the case of a strong enough magnetism of the F layer a superconducting
phase difference on the junction electrodes is about $\pi /2$. If such a
contact is in magnetic field, an additional phase shift on the electrodes
has been induced, and, depending on the relative orientation, the applied
field prompts the junction to the $\pi $ - or to the $0$ - phase state. From
the fundamental point of view, these results provide new effects of the
superconductivity in the presence of a spin splitting in hybrid
superconductor-ferromagnet-normal metal structures and may lead to further
understanding of the proximity effects in ferromagnetic materials. The
feature important for practical applications is that the superconducting
properties of S/F-I-N/S junctions can be simply varied by changing the
external magnetic field direction.

Some simplifying assumptions have been made, such as the model admitted of
analytical solutions. We consider the dirty limit. However, when the
thickness of the F and N layers is much less then its coherence length, $%
d_{F(N)}<<\xi _{F(N)}$ , the results are, in most cases, applicable for
clean F and N metals, too. Indeed, if scattering centers are on average
separated by a distance smaller than the coherence length, then the dirty
limit may be applied. The physical presence of interfaces with atomic
roughness ensures that there are such scattering processes for very thin
layers.

Let us estimate the orbital effects. The magnetic field H induces screening
currents and leads to some suppression of the condensate function. For
ferromagnetic and nonmagnetic layers in the dirty limit the depairing rate
due to the Meissner current is determined by the energy $Dp_{S}^{2}$ , where 
$p_{S}\approx Hd_{F(N)}/\phi _{O}$ is the condensate momentum (here $%
Hd_{F(N)}$ is the magnitude of the vector potential into the F (N) layer and 
$\phi _{O}$ is the flux quantum). So, the orbital effects can be neglected
if 
\[
Dp_{S}^{2}\symbol{126}(D/\xi _{F(N)}^{2})(H/H_{C2})^{2}(d_{F(N)}/\xi
_{F(N)})^{2}\symbol{126}\Delta _{S}(H/H_{C2})^{2}(d_{F(N)}/\xi
_{F(N)})^{2}<<H,H_{e}\text{ \ , } 
\]
where $H_{C2}$ \symbol{126}$\phi _{O}/\xi _{F(N)}^{2}$ is the upper critical
field of F (N) layer. We see, that the nanoscale nonsuperconducting layers
thickness, $d_{F(N)}<<\xi _{F(N)}$ , ensures that there is a fairly uniform
penetration of an applied field and that orbital depairing effects are
minimized. However, for nanoscale hybrid structures a strong electric field
arising near metal-metal boundaries may be a source of spin-flip processes
[21]. Decreasing the average (ferro)magnetic field experienced by the Cooper
pair, the spin-flip scattering is much more destructive than the orbital
effect [22,23].

It was also assumed that the applied field is concentrated into the ''weak''
F-I-N part of the junction. In a real situation, an external magnetic field
can partly penetrate into the S layers and cause some suppression of the
order parameter $\Delta _{S}$ . This issue was already addressed a while ago
and well understood (see, e.g., de Gennes [24], Chap.VII). While, in the
general case, it is necessary to take into account self-consistently
corrections to the pair potential of the S layer, this fact, however, does
not qualitatively influence the results of the paper, if the magnetic flux
that penetrates into the contact is much less that a flux quantum. The
condition is fulfilled when the transverse dimensions of the junction are
much less then the Josephson penetration depth.

Let us briefly comment on an experimental set-up, as well. Several recent
works (see, e. g., [5-7,25]) have found that dilute ferromagnetic $%
Cu_{x}Ni_{1-x}$ alloys (with x 
\mbox{$<$}%
0.6) are most adopted as materials for the F layers, since their weak
ferromagnetism is less devastating to superconductivity. Magnetic hysteresis
measurements show for $Cu_{x}Ni_{1-x}$ films with x $\approx $ 0.4$\div $0.5
the coercive field of \symbol{126}100 Oe (see, e.g., [25] and references
therein ). One can also pin the magnetization of the $CuNi$ layer by
adjacent $FeMn$ layer via exchange bias, and in such a way enlarge the
interval of an applied magnetic field, where the magnetization remains
fixed. On the other hand, various niobium oxides, one of them NbO, possess
the properties of a normal nonmagnetic metal. Such an N layer is usually
formed during the fabrication process (see, e. g., [26] and references
therein).\ So, conventional, e. g., magnetron sputtering technique can be
used to grow the tunnel superconductor-ferromagnet-insulator-nonmagnetic
metal-superconductor structures.

The authors are grateful to M. A. Belogolovskii for reading the paper and
useful discussions.

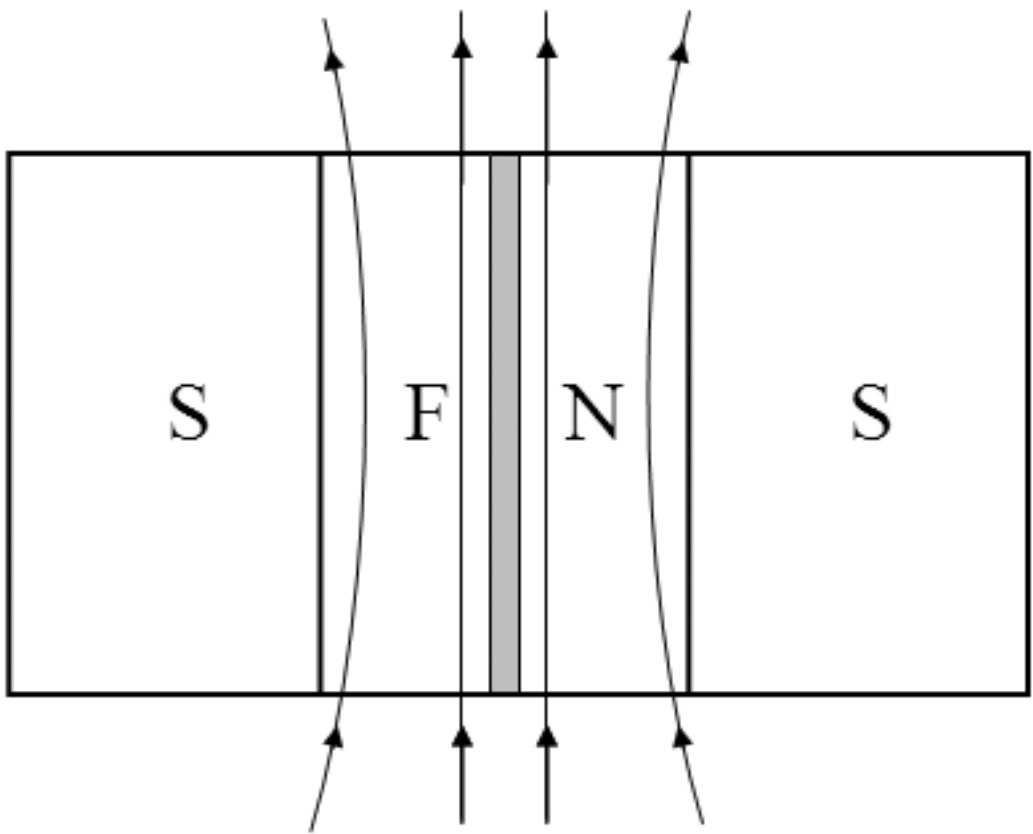
Fig. 1. The S/F-I-N/S system in a parallel magnetic field.

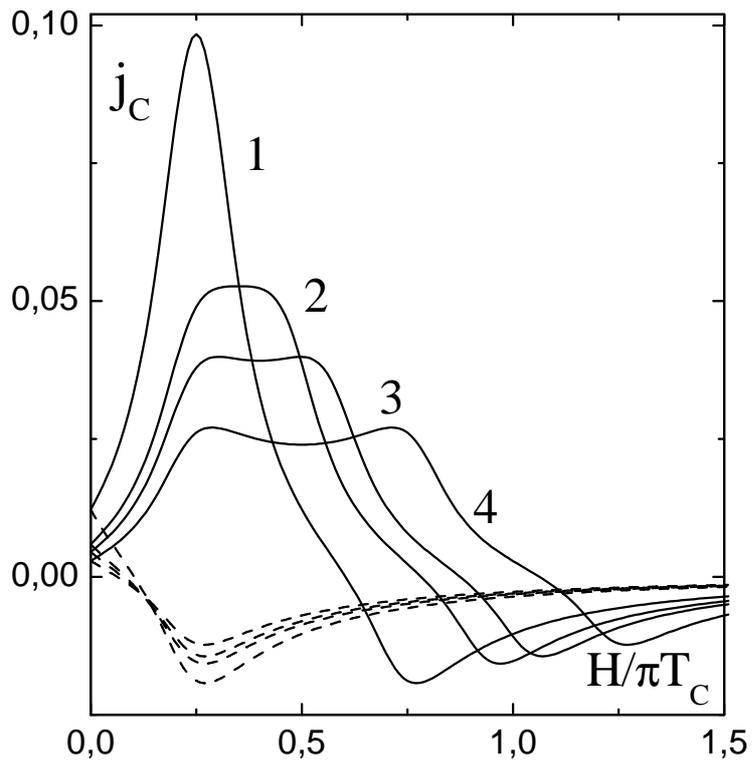
Fig. 2.\ Critical current of the S/F-I-N/S tunnel junction vs external
magnetic field at high interface transparency, $\gamma _{B}=0$ , and strong
proximity effect, $\gamma _{M}$ = 10, for different values of the exchange
field of the F layer: $H_{e}/\pi T_{C}=$ 0.5, 0.7, 0.8 and 1.0 (curves 1, 2,
3 and 4, respectively). Here and on Figs. 3-7, solid curves illustrate the
case of antiparallel orientation of the internal F layer exchange field and
the external magnetic field; dashed curves illustrate the case of parallel
orientation of these fields. The numerical results where obtained for $%
T=0.1T_{C}$ .

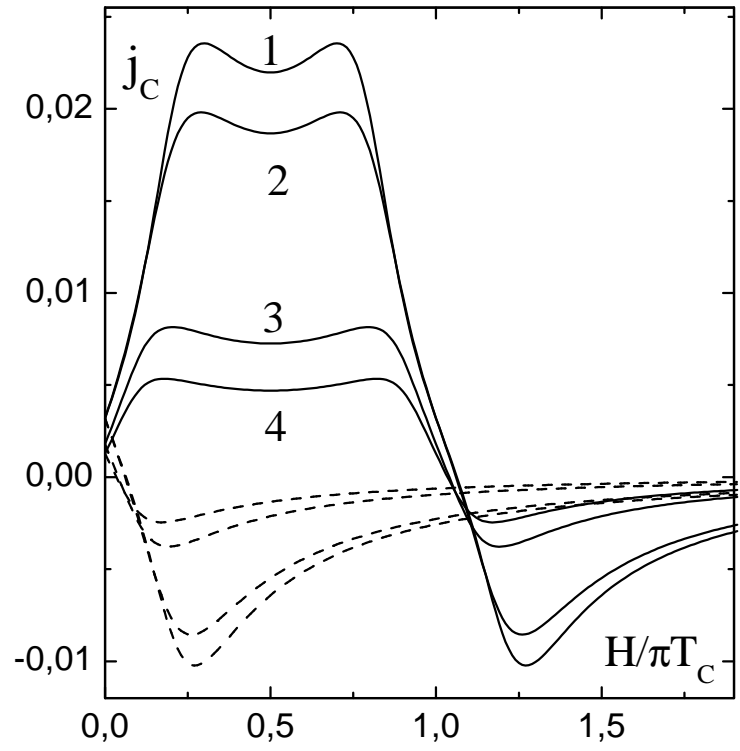
Fig. 3. Critical current of the S/F-I-N/S tunnel junction vs external
magnetic field at strong proximity effect, $\gamma _{M}=10$, $H_{e}=\pi
T_{C} $\ and different values of SF (SN) boundary transparency: $\gamma
_{B}= $ 1, 2, 7, and 10 (curves 1, 2, 3 and 4, respectively).

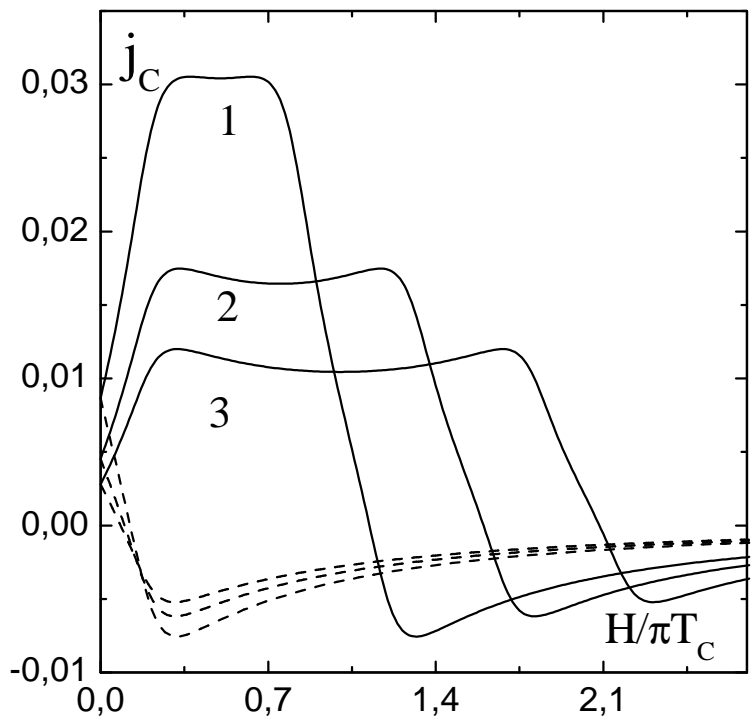
Fig. 4.\ Critical current vs external magnetic field for the S/F-I-N/S
tunnel junction with $\gamma _{B}=5$ and weak proximity effect $\gamma _{M}$
= 0.1 for different values of the exchange field of the F layer: $H_{e}/\pi
T_{C}=$ 1.0, 1.5, and 2.0 (curves 1, 2, and 3, respectively).

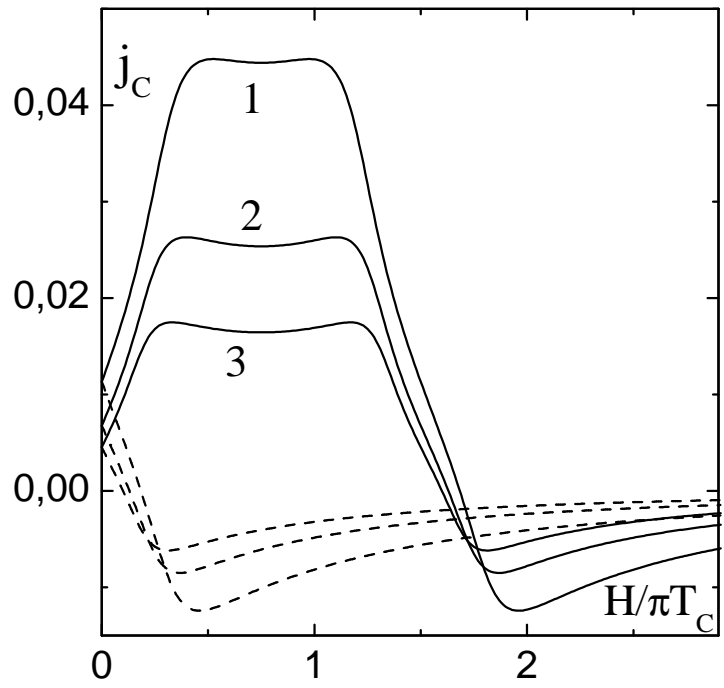
Fig. 5. Critical current of an S/F-I-N/S tunnel junction vs exchange energy
and weak proximity effect $\gamma _{M}=0.1$, $H_{e}=1.5\pi T_{C}$\ and
different values of SF (SN) boundary transparency: $\gamma _{B}=$ 3, 4, and
5 (curves 1, 2, and 3, respectively).

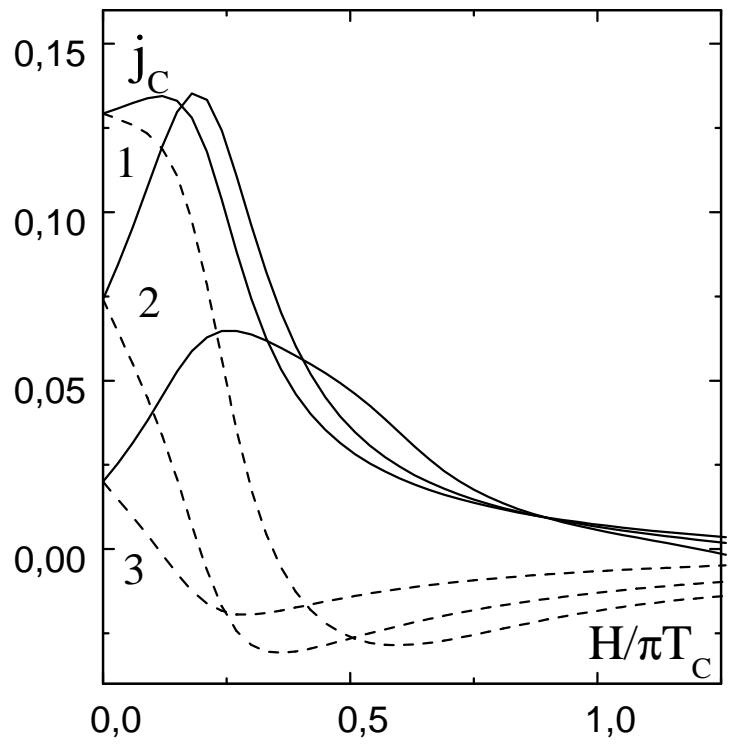
Fig. 6. Critical current for nonsymmetric S/F-I-N/S tunnel structure with a
strong proximity effect for the SN boundary, $\gamma _{MN}=10$ , and a weak
proximity effect for the SF boundary, $\gamma _{MF}=0.1$ , $H_{e}=\pi T_{C}$%
\ and different values of SF (SN) interface transparency: $\gamma _{B}=$
0.5, 1, and 2 (curves 1, 2 and 3, respectively).

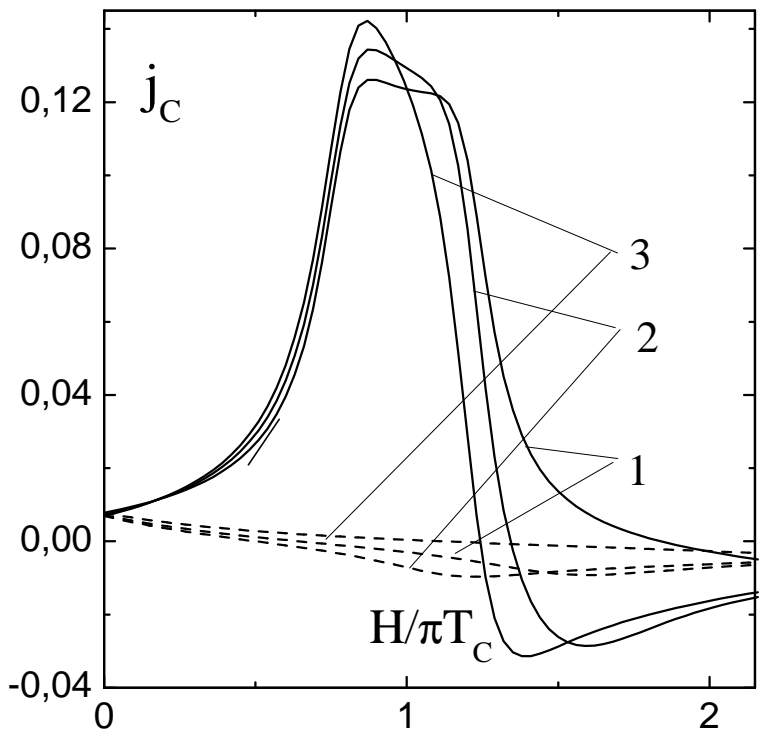
Fig. 7.\ Critical current of nonsymmetric S/F-I-N/S tunnel structure with a
strong proximity effect for the SF boundary, $\gamma _{MF}=10$ , and a weak
proximity effect for the SN boundary, $\gamma _{MN}=0.1$ , $H_{e}=\pi T_{C}$%
\ and high values of SF (SN) interface transparency: $\gamma _{B}=$ 0.2,
0.5, 0.7 (curves 1, 2 and 3, respectively).

Fig. 8 The dependence of the critical current on the angle $\theta $\
between the fields for S/F-I-N/S structure with $\gamma _{M}=0.1$, $\gamma
_{B}=5\ $for SF and SN boundaries and$\ H_{e}=\pi T_{C}$\ ; $\theta $\ $%
=0^{o},30^{o},60^{o},90^{o},120^{o},150^{o},180^{o}$\ (curves 1 -- 7,
respectively).

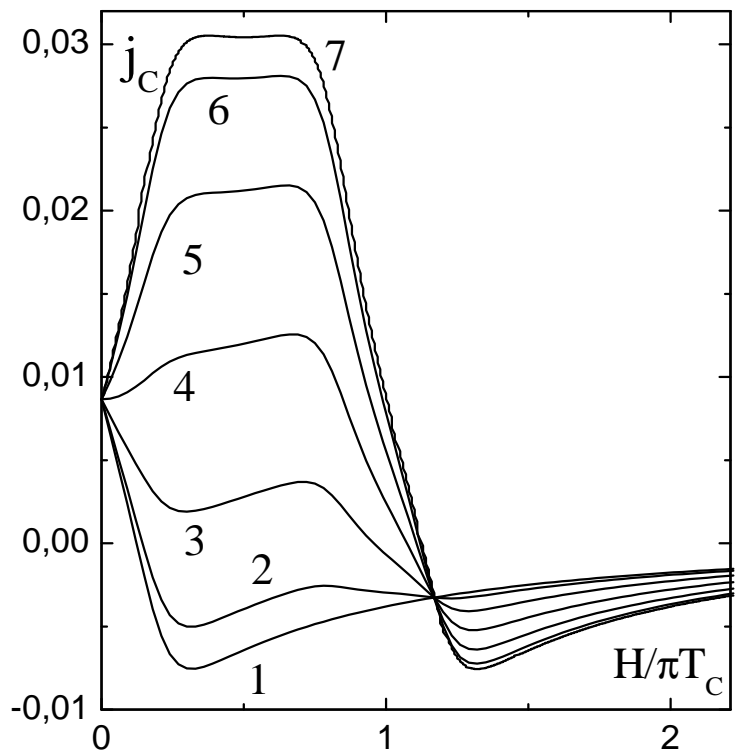
\end{document}